\documentclass[conference]{IEEEtran}
\IEEEoverridecommandlockouts

\usepackage{amsmath,amssymb,amsfonts}
\usepackage{graphicx}
\usepackage{booktabs}
\usepackage{cite}
\usepackage{algorithm}
\usepackage{algorithmic}
\usepackage{url}

\title{When Discovery Outpaces Remediation: Modeling AI-Accelerated Vulnerability Discovery in Interconnected Systems}

\author{
\IEEEauthorblockN{Mohamamad Reza Faghani, PhD}
\IEEEauthorblockA{York University\\
Email: mrfaghani@gmail.com}
}

\begin{document}
\maketitle

\begin{abstract}
Advanced AI systems for code analysis, binary analysis, fuzzing orchestration, and penetration-test planning may significantly increase the rate at which latent vulnerabilities are discovered. While improved discovery can benefit defenders, it can also overload remediation pipelines and accelerate adversarial weaponization. This paper develops a queueing and network-theoretic model of AI-accelerated vulnerability discovery in interconnected systems. We represent an enterprise as a weighted dependency graph with replenishing vulnerability pools, finite remediation capacity, triage degradation, exploit-window compression, and dynamic compromise propagation. We derive stability conditions for vulnerability backlogs, formulate a dynamic coupling between unresolved backlog and cascade risk, and evaluate mitigation strategies through simulation. Results indicate that when actionable discovery arrivals exceed remediation throughput, backlogs grow rapidly and systemic risk increases nonlinearly. In hub-dominated topologies, segmentation can reduce propagated compromise more effectively than remediation speed alone, while the strongest defense combines remediation automation with reduced network coupling.
\end{abstract}


\section{Introduction}
Modern organizations defend ecosystems of cloud services, endpoints, identity providers, CI/CD pipelines, APIs, operational technology, and third-party integrations rather than isolated hosts. Security outcomes are determined not only by the vulnerabilities present on individual assets, but also by how systems are connected, how privileges propagate, and how quickly discovered weaknesses can be remediated.

Traditional vulnerability management implicitly assumes that vulnerability discovery is scarce. Human researchers, red teams, scanners, and bug bounty programs surface only a fraction of latent flaws over any finite horizon. Under this assumption, improved discovery generally improves security because more vulnerabilities become known and can be patched before exploitation.

Advanced AI systems may invalidate this assumption. AI-assisted code review, binary analysis, fuzzing orchestration, symbolic execution, exploit reasoning, and automated testing can increase the rate at which latent vulnerabilities become known. If vulnerabilities are found faster than organizations can validate, prioritize, patch, test, deploy, and verify fixes, then cybersecurity shifts from a search-constrained discipline to a remediation-capacity discipline.

This paper studies that transition. We model an enterprise as a weighted dependency graph with replenishing latent vulnerability pools, finite remediation queues, triage degradation under overload, exploit-window compression, and dynamic compromise propagation across trust edges. The central question is not simply whether AI discovers more vulnerabilities, but whether the resulting actionable arrival stream exceeds the organization's capacity to absorb and contain it.

We make five contributions:
\begin{enumerate}
    \item We introduce a replenishing-pool vulnerability model suitable for continuously deployed systems.
    \item We formulate vulnerability remediation as a queueing process and derive local and long-run stability conditions.
    \item We model triage degradation and exploit-window compression as mechanisms that convert backlog into exposure.
    \item We develop a dynamic cascade model in which unresolved backlog directly affects compromise hazard and propagation susceptibility.
    \item We provide a simulation study comparing baseline operation, automated remediation, segmentation, and combined defenses across random, small-world, and scale-free networks.
\end{enumerate}

Our results suggest that discovery acceleration alone is insufficient as a defensive strategy. We also outline practical pathways for calibrating the model from operational telemetry. In high-discovery-rate environments, resilience depends on remediation throughput, exploit-aware prioritization, attack-surface reduction, and reduced coupling among systems.

\section{Related Work}
\subsection{Vulnerability Discovery Models}
Software reliability growth models and vulnerability discovery models often represent discovered flaws as realizations of a counting process over a finite latent pool. Classical non-homogeneous Poisson process models assume that the discovery rate is proportional to the remaining undiscovered defects or vulnerabilities \cite{goel1979,yamada1984,lyu1996,pham2006}. Security-specific vulnerability discovery models extend this logic to vulnerability disclosure and discovery data \cite{ozment2007,ozment2006,rescorla2005,frei2006,alhazmi2007}. These models are useful for stable software systems but are less suited to continuously deployed environments in which new vulnerabilities are introduced through frequent releases, dependency updates, and configuration drift.

Our work differs by explicitly modeling replenishment of the latent vulnerability pool. This avoids treating modern software as a one-time finite depletion process and allows analysis of long-run queue stability.

\subsection{Queueing and Remediation Capacity}
Enterprise patch management is constrained by ownership boundaries, change windows, regression risk, testing requirements, vendor dependencies, and operational priorities \cite{nist80040}. Queueing theory provides the natural language for such systems: vulnerability findings arrive as work, while remediation teams and automated deployment pipelines provide finite service capacity \cite{kleinrock1975,harchol2013}. This motivates our queueing abstraction and the long-run stability condition $\bar a_i<\bar\mu_i$.

\subsection{Attack Graphs and Network Propagation}
Attack graphs model how vulnerabilities can be chained across systems to reach high-value targets \cite{sheyner2002,noel2004,ou2006}. Network science and epidemic models have also been used to study malware propagation, cascading failure, and systemic risk in complex networks \cite{newman2010,pastor2001,pastor2015,wang2003,vanmieghem2009,ganesh2005}. Spectral-radius thresholds are common in epidemic processes on networks and motivate our cascade criticality theorem.

Cyber-malware propagation on structured networks has also been studied in the context of XSS worms in online social networks, where clustering, community structure, and strategically selected monitoring nodes significantly affect spread and detection performance \cite{xssworm2013}.

\subsection{AI for Cybersecurity}
AI has been applied to malware detection, anomaly detection, static analysis, vulnerability prediction, fuzzing, and security operations. Recent large-model systems can also assist with code reasoning, tool orchestration, test generation, and penetration-test planning. The dominant framing is that AI improves discovery and analysis. We focus instead on the system-level consequence: discovery acceleration may shift the security bottleneck from finding vulnerabilities to remediating and containing them.

\section{System Model}
\subsection{Enterprise Dependency Graph}
We represent an enterprise as a directed weighted graph:
\begin{equation}
G=(N,E,W),
\end{equation}
where $N=\{1,2,\ldots,n\}$ is the set of systems, $E\subseteq N\times N$ is the set of directed trust or dependency relationships, and $W=[w_{ij}]$ is a matrix of edge weights. An edge $(i,j)$ indicates that compromise or failure at node $i$ can affect node $j$ through network reachability, shared identity, API trust, operational dependency, administrative privilege, or data access.

The weight $w_{ij}\in[0,1]$ represents propagation strength. High-weight edges include shared administrative credentials, unrestricted network reachability, implicit service trust, CI/CD deployment rights, and centralized identity privileges.

\subsection{Node State Variables}
Each node $i$ has attack surface $A_i$, latent vulnerabilities $U_i(t)$, known unresolved backlog $B_i(t)$, remediation capacity $\mu_i(t)$, exploitability factor $E_i(t)$, business criticality $C_i$, and segmentation/control strength $S_{ij}$ on incident edges.

Attack surface $A_i$ may be approximated by exposed services, reachable interfaces, dependency count, code complexity, privilege level, and configuration complexity. Business criticality $C_i$ measures consequence of compromise.

\subsection{Replenishing Vulnerability Pool}
Continuously deployed systems are not finite depletion systems. Vulnerabilities are introduced through new code, dependency upgrades, configuration drift, changing execution environments, and operational mistakes. We model the latent pool as:
\begin{equation}
\frac{dU_i(t)}{dt}=\eta_i(t)-d_i(t)-\chi_i(t),
\label{eq:latent}
\end{equation}
where $\eta_i(t)$ is the vulnerability introduction rate, $d_i(t)$ is the discovery rate from the latent pool, and $\chi_i(t)$ is removal without first entering backlog, such as refactoring, decommissioning, hardening, or architectural simplification.

This distinction is essential. If $\eta_i(t)=0$, vulnerability discovery eventually depletes a finite pool. If $\eta_i(t)>0$, the system can experience persistent vulnerability pressure even after early latent flaws are discovered.

\section{Discovery and Queueing Dynamics}
\subsection{Discovery Intensity}
Let baseline discovery intensity be:
\begin{equation}
\lambda_i(t)=\alpha_i A_i q_i(t),
\end{equation}
where $\alpha_i$ is a discovery-effort coefficient and $q_i(t)$ captures scanner coverage, researcher attention, test quality, and observability. AI acceleration is represented by $k_i(t)\geq 1$:
\begin{equation}
\lambda'_i(t)=k_i(t)\lambda_i(t).
\end{equation}

The rate at which latent vulnerabilities become known is:
\begin{equation}
d_i(t)=\lambda'_i(t)U_i(t).
\end{equation}

Not all discoveries are valid, exploitable, or actionable. Let $\theta_i\in[0,1]$ denote the actionable fraction. Then actionable arrivals to the remediation queue are:
\begin{equation}
a_i(t)=\theta_i\lambda'_i(t)U_i(t).
\end{equation}

\subsection{Backlog Dynamics}
Let $B_i(t)$ denote unresolved actionable vulnerabilities. Remediation is modeled as finite service capacity:
\begin{equation}
\frac{dB_i(t)}{dt}=a_i(t)-r_i(t),
\end{equation}
where
\begin{equation}
r_i(t)=\min(B_i(t),\mu_i(t)).
\end{equation}

\subsection{Local and Long-Run Stability}
For a short quasi-stationary interval in which $U_i(t)\approx U_i$, backlog stability requires:
\begin{equation}
\theta_i k_i\alpha_iA_iq_iU_i < \mu_i.
\end{equation}
The corresponding local critical acceleration factor is:
\begin{equation}
k_{i,local}^{crit}=\frac{\mu_i}{\theta_i\alpha_iA_iq_iU_i}.
\end{equation}

For continuously deployed systems, the relevant condition is long-run:
\begin{equation}
\bar{a}_i < \bar{\mu}_i,
\end{equation}
where
\begin{equation}
\bar{a}_i=\limsup_{T\rightarrow\infty}\frac{1}{T}\int_0^T \theta_i\lambda'_i(t)U_i(t)dt.
\end{equation}

This resolves the distinction between two regimes. AI acceleration can create a transient discovery shock by draining a large latent pool, while continuous deployment creates a persistent arrival floor determined by vulnerability introduction and removal dynamics.

\section{Triage Degradation and Exploit-Window Compression}
\subsection{Triage Under Overload}
Discovery acceleration does not only increase the number of remediation tickets. It can also degrade prioritization quality by overwhelming validation, ownership assignment, and risk-ranking processes. Let $H(t)$ denote triage capacity and $A(t)=\sum_i a_i(t)$ denote actionable arrivals. Define triage load:
\begin{equation}
L_T(t)=\frac{A(t)}{H(t)}.
\end{equation}

We model triage quality as:
\begin{equation}
Q_T(t)=\frac{1}{1+\gamma(L_T(t)-1)^+},
\end{equation}
where $(x)^+=\max(x,0)$ and $\gamma$ controls overload sensitivity. When $L_T(t)\leq1$, triage quality remains near one. When $L_T(t)>1$, high-risk vulnerabilities may be delayed behind lower-risk findings, increasing exposure.

\subsection{Exploit-Window Compression}
Let $\delta_v$ be the time between vulnerability discovery and reliable exploit availability. If adversaries also benefit from AI-assisted analysis, exploit development time may compress:
\begin{equation}
\delta'_v=\frac{\delta_v}{m_v}, \quad m_v\geq1.
\end{equation}

Let $T_{rem,v}$ be remediation time. The exploitable exposure window is:
\begin{equation}
\Omega_v(t)=\max(0,T_{rem,v}-\delta'_v).
\end{equation}

Exposure load should therefore increase with backlog, exploitability, criticality, triage degradation, and exploit acceleration. We use:
\begin{equation}
X_i(t)=\sum_{v\in B_i(t)} \frac{E_{iv}C_iR_i\Omega_v(t)}{Q_T(t)}.
\end{equation}
This formulation restores an important mechanism: discovery overload increases risk not merely by adding tickets, but by reducing the quality and timeliness of defensive prioritization.

\section{Dynamic Cascade Model}
\subsection{Backlog-Driven Exposure}
Unresolved vulnerabilities increase compromise exposure. The exposure load $X_i(t)$ from Eq. (16) determines local compromise hazard:
\begin{equation}
h_i(t)=\beta X_i(t),
\end{equation}
where $\beta$ maps exposure load to compromise hazard.

\subsection{Network Propagation}
Let $c_i(t)\in[0,1]$ denote the compromise probability or compromise state of node $i$. Network hazard into node $j$ is:
\begin{equation}
h_j^{net}(t)=\sum_{i\in \mathcal{N}^{-}(j)}c_i(t)w_{ij}(1-S_{ij})\psi_j(B_j(t)),
\end{equation}
where $S_{ij}\in[0,1]$ is segmentation strength and $\psi_j(B_j(t))$ captures susceptibility induced by unresolved backlog. We use:
\begin{equation}
\psi_j(B_j(t))=1-\exp[-\zeta B_j(t)].
\end{equation}

The total hazard is:
\begin{equation}
h_j^{tot}(t)=h_j(t)+h_j^{net}(t).
\end{equation}

Compromise evolves as:
\begin{equation}
c_j(t+\Delta t)=1-(1-c_j(t))\exp[-h_j^{tot}(t)\Delta t].
\end{equation}

This formulation couples queue state and cascade dynamics: discovery increases backlog, backlog increases susceptibility, compromised neighbors add network hazard, and segmentation reduces edge-mediated propagation.

\subsection{Quasi-Static Cascade Criticality}
The cascade layer requires an analytical stability condition rather than only simulation. Because the backlog vector $B(t)$ evolves over time, the threshold below is a quasi-static condition: it applies on intervals where the cascade evolves faster than the backlog, or where $B(t)$ is treated as frozen for local stability analysis. This is analogous to threshold analysis in time-varying epidemic systems, where an instantaneous reproduction condition is informative when parameter drift is slow relative to transmission dynamics.

We derive the threshold by linearizing the propagation dynamics around the disease-free state $c_i=0$. Let
\begin{equation}
M_{ij}=w_{ij}(1-S_{ij})\psi_j(B_j)
\end{equation}
be the effective propagation matrix for fixed backlog state $B$. Let $\rho(M)$ denote its spectral radius. Assume a compromised node remains infectious for mean duration $1/\delta_c$, where $\delta_c>0$ is the recovery, containment, or isolation rate. Let $\kappa$ scale global propagation strength. Then the linearized dynamics are:
\begin{equation}
\dot{c}(t) \approx (\kappa M^{T}-\delta_c I)c(t).
\end{equation}

\textbf{Theorem 1 (Quasi-Static Backlog-Coupled Cascade Threshold).}
For fixed backlog vector $B$ and effective propagation matrix $M$, the disease-free equilibrium is locally stable if
\begin{equation}
\kappa \rho(M) < \delta_c,
\end{equation}
and unstable if
\begin{equation}
\kappa \rho(M) > \delta_c.
\end{equation}
Equivalently, the critical coupling threshold is:
\begin{equation}
\kappa_c(B)=\frac{\delta_c}{\rho(M)}.
\end{equation}

\textit{Proof.}
The linearized cascade dynamics are governed by the Jacobian $J=\kappa M^T-\delta_c I$. The disease-free equilibrium is locally asymptotically stable if every eigenvalue of $J$ has negative real part. Since $M$ is nonnegative, the Perron--Frobenius theorem implies that its dominant eigenvalue is real and equal to $\rho(M)$. The dominant eigenvalue of $J$ is therefore $\kappa\rho(M)-\delta_c$. Stability requires $\kappa\rho(M)-\delta_c<0$, yielding $\kappa\rho(M)<\delta_c$. Instability follows when this quantity is positive. Solving the equality gives $\kappa_c(B)=\delta_c/\rho(M)$. \hfill $\square$

The theorem should not be interpreted as a global stability guarantee for arbitrary, rapidly changing $B(t)$. If backlog changes on the same timescale as propagation, $
ho(M(t))$ becomes time-varying and the relevant condition is path-dependent. Nevertheless, the quasi-static threshold is useful because it identifies the instantaneous cascade margin and explains how remediation and segmentation affect that margin.

This theorem links the queue layer and cascade layer. Since $\psi_j(B_j)=1-\exp[-\zeta B_j]$ is nondecreasing in $B_j$, unresolved backlog increases entries of $M$, weakly increases $\rho(M)$, and therefore weakly decreases $\kappa_c(B)$. In other words, queue overload lowers the amount of network coupling required for a self-sustaining cascade.

\subsection{Systemic Risk}
We define criticality-weighted systemic risk as:
\begin{equation}
\mathcal{R}(t)=\sum_{i=1}^{n}C_ic_i(t).
\end{equation}

The statement that increasing $S_{ij}$ reduces propagation follows immediately from the nonnegative edge term and is treated as a monotonicity property rather than a central theorem. The substantive analytical result is Theorem 1: segmentation changes the spectral radius of the effective propagation matrix and can move the system below the cascade threshold.

\section{Simulation Study}
\subsection{Design}
We implement a discrete-time Monte Carlo simulation of vulnerability introduction, discovery, triage, remediation, exploit-window compression, and propagation. This version updates the simulation to match the dynamic cascade model in Section V: backlog affects both local hazard and downstream susceptibility through $\psi_j(B_j(t))$.

Each run contains $n=260$ systems over $T=70$ time steps. We report averages over 24 trials per topology and acceleration factor, and 32 trials for the intervention analysis. Figures report 95\% confidence intervals. We evaluate random, small-world, and scale-free topologies.

The AI discovery acceleration factor is varied over:
\begin{equation}
k\in\{1,2,3,4,6,8,10\}.
\end{equation}

Each node is assigned heterogeneous attack surface, initial latent vulnerabilities, remediation capacity, exploitability, and criticality. Each edge is assigned a propagation weight. At each time step, latent vulnerabilities are introduced, AI-accelerated discovery surfaces vulnerabilities, actionable findings enter backlog, remediation removes findings up to capacity, triage quality is updated as a function of load, and unresolved exposure contributes to local and propagated compromise probability.

The backlog susceptibility parameter $\zeta$ controls how quickly unresolved backlog increases propagation susceptibility. Because $\zeta$ is not directly observable in most organizations, we treat it as a sensitivity parameter rather than a calibrated constant. Unless otherwise stated, simulations use $\zeta=0.55$ and Section VII-C reports sensitivity over $\zeta\in[0.05,2.0]$.

\subsection{Algorithm}
\begin{algorithm}[htbp]
\caption{Backlog-Coupled Vulnerability Cascade Simulation}
\begin{algorithmic}[1]
\STATE Generate graph $G=(N,E,W)$
\STATE Initialize $A_i,U_i,B_i,C_i,\mu_i,c_i$ for each node
\FOR{$t=1$ to $T$}
    \FOR{each node $i$}
        \STATE Introduce latent vulnerabilities into $U_i$
        \STATE Remove latent vulnerabilities through hardening/refactoring
        \STATE Discover vulnerabilities at rate $k\alpha_iA_iq_iU_i$
        \STATE Add actionable discoveries to $B_i$
        \STATE Remediate up to capacity $\mu_i$
    \ENDFOR
    \STATE Compute triage load $L_T(t)$ and quality $Q_T(t)$
    \FOR{each node $i$}
        \STATE Compute exposure load $X_i(t)$ with exploit-window compression
        \STATE Compute local hazard $h_i(t)$
    \ENDFOR
    \FOR{each node $j$}
        \STATE Compute susceptibility $\psi_j(B_j(t))$
        \STATE Compute network hazard $h_j^{net}(t)$
        \STATE Update compromise probability $c_j(t)$
    \ENDFOR
    \STATE Record backlog, systemic risk, time-to-critical compromise, and cascade size
\ENDFOR
\end{algorithmic}
\end{algorithm}

\subsection{Intervention Scenarios}
We compare four scenarios:
\begin{enumerate}
    \item Baseline: no additional intervention.
    \item Automated remediation: remediation capacity is doubled.
    \item Segmentation: edge propagation strength is reduced by 60\%.
    \item Combined: remediation capacity is doubled and segmentation is applied.
\end{enumerate}

\section{Results}
\subsection{Backlog Growth}
\begin{figure}[htbp]
\centerline{\includegraphics[width=0.48\textwidth]{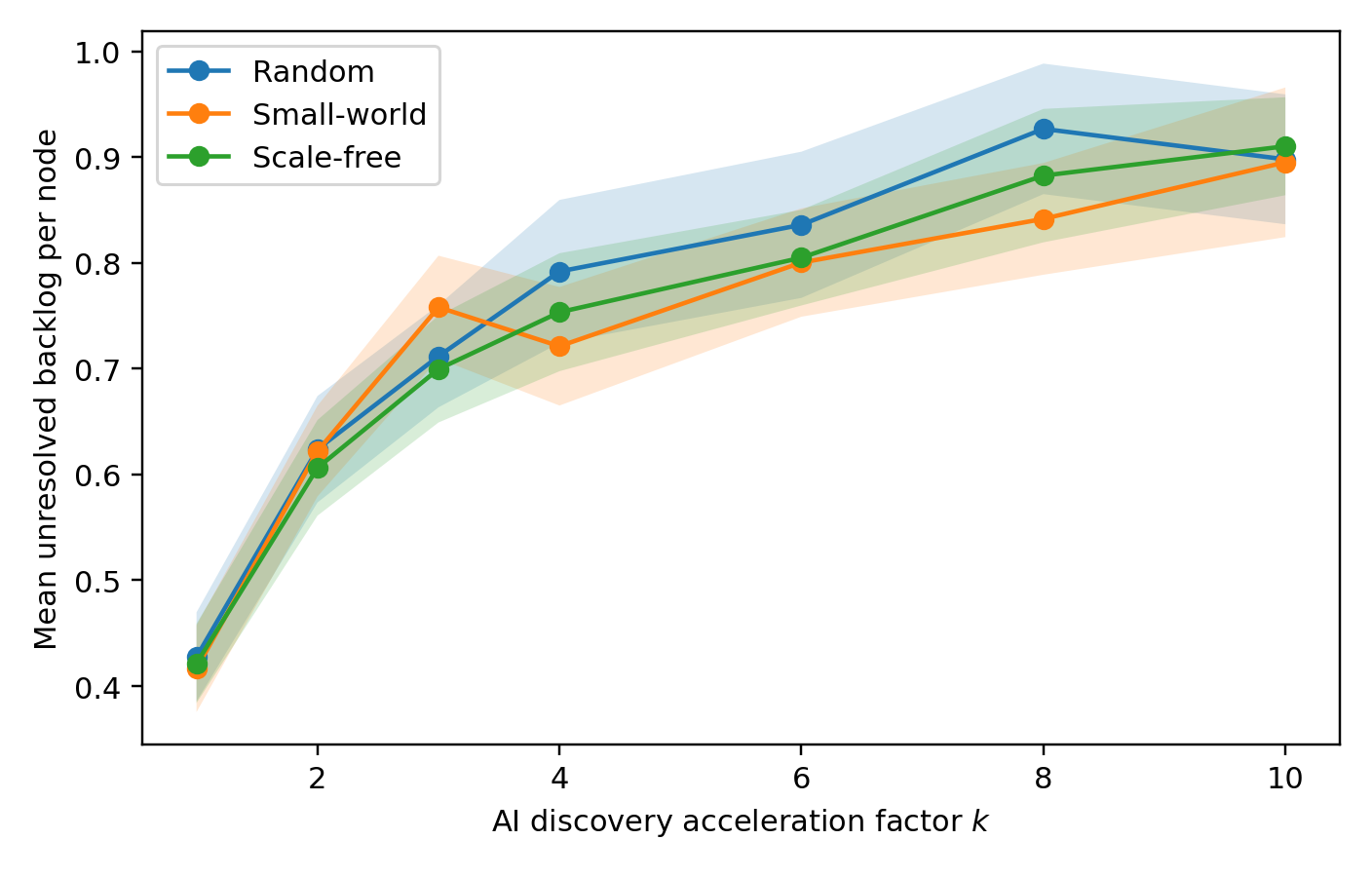}}
\caption{Mean unresolved backlog per node as AI discovery acceleration increases. Shaded bands denote 95\% confidence intervals.}
\label{fig:backlog}
\end{figure}

Fig.~\ref{fig:backlog} shows that unresolved backlog grows with AI discovery acceleration. Once actionable arrivals exceed remediation capacity, unresolved work accumulates rapidly. Shaded bands denote 95\% confidence intervals across simulation trials.

\subsection{Systemic Risk}
\begin{figure}[htbp]
\centerline{\includegraphics[width=0.48\textwidth]{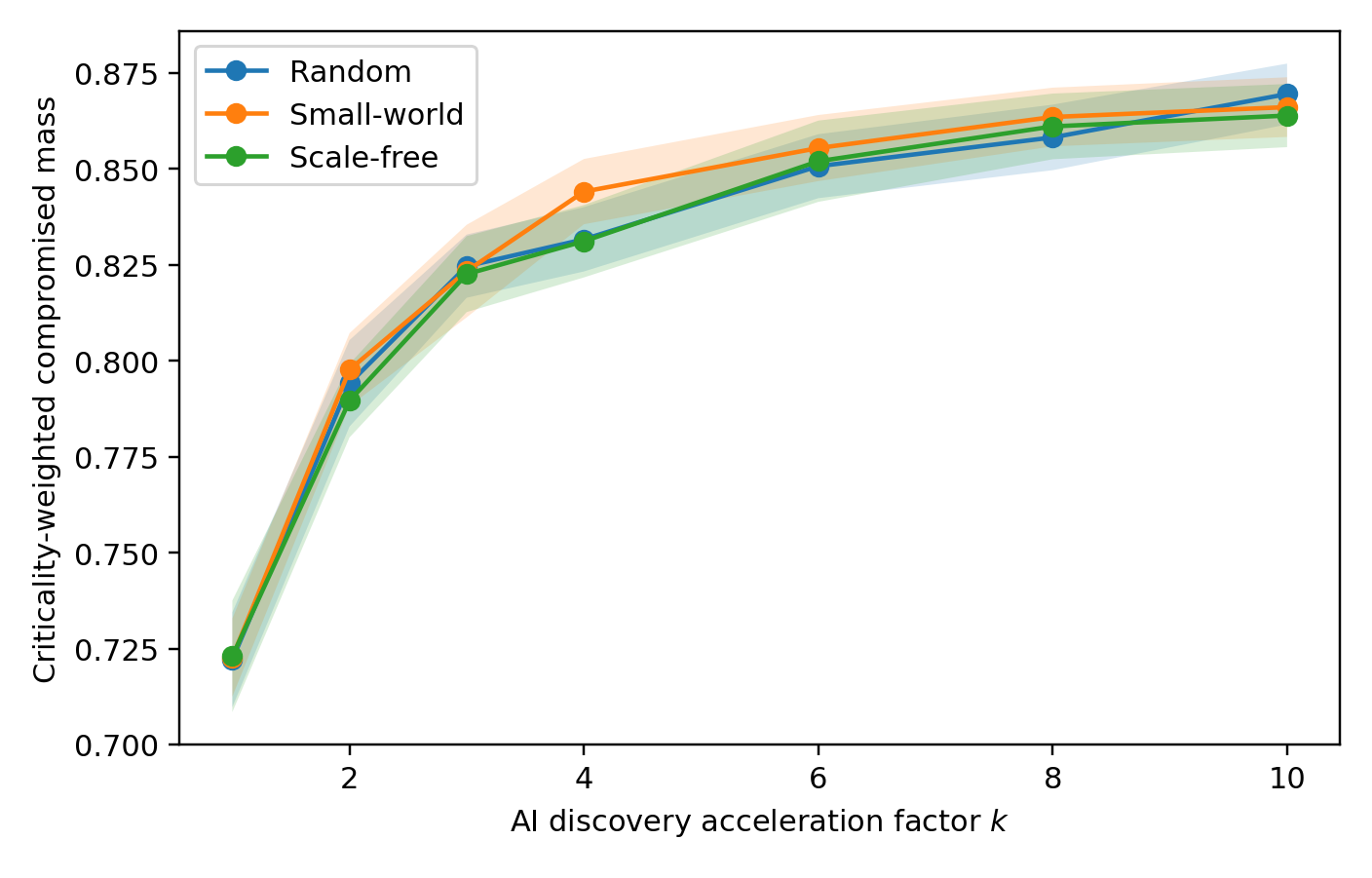}}
\caption{Criticality-weighted compromised mass at the end of the simulation. Shaded bands denote 95\% confidence intervals.}
\label{fig:sysrisk}
\end{figure}

Fig.~\ref{fig:sysrisk} shows that criticality-weighted compromised mass increases with discovery acceleration. Small-world and scale-free graphs exhibit higher systemic risk than random graphs because short paths and hubs amplify propagation.

This aligns with prior findings that clustering and community boundaries can materially alter malicious propagation dynamics in structured networks \cite{xssworm2013}.

\subsection{Time-to-Critical Compromise}
\begin{figure}[htbp]
\centerline{\includegraphics[width=0.48\textwidth]{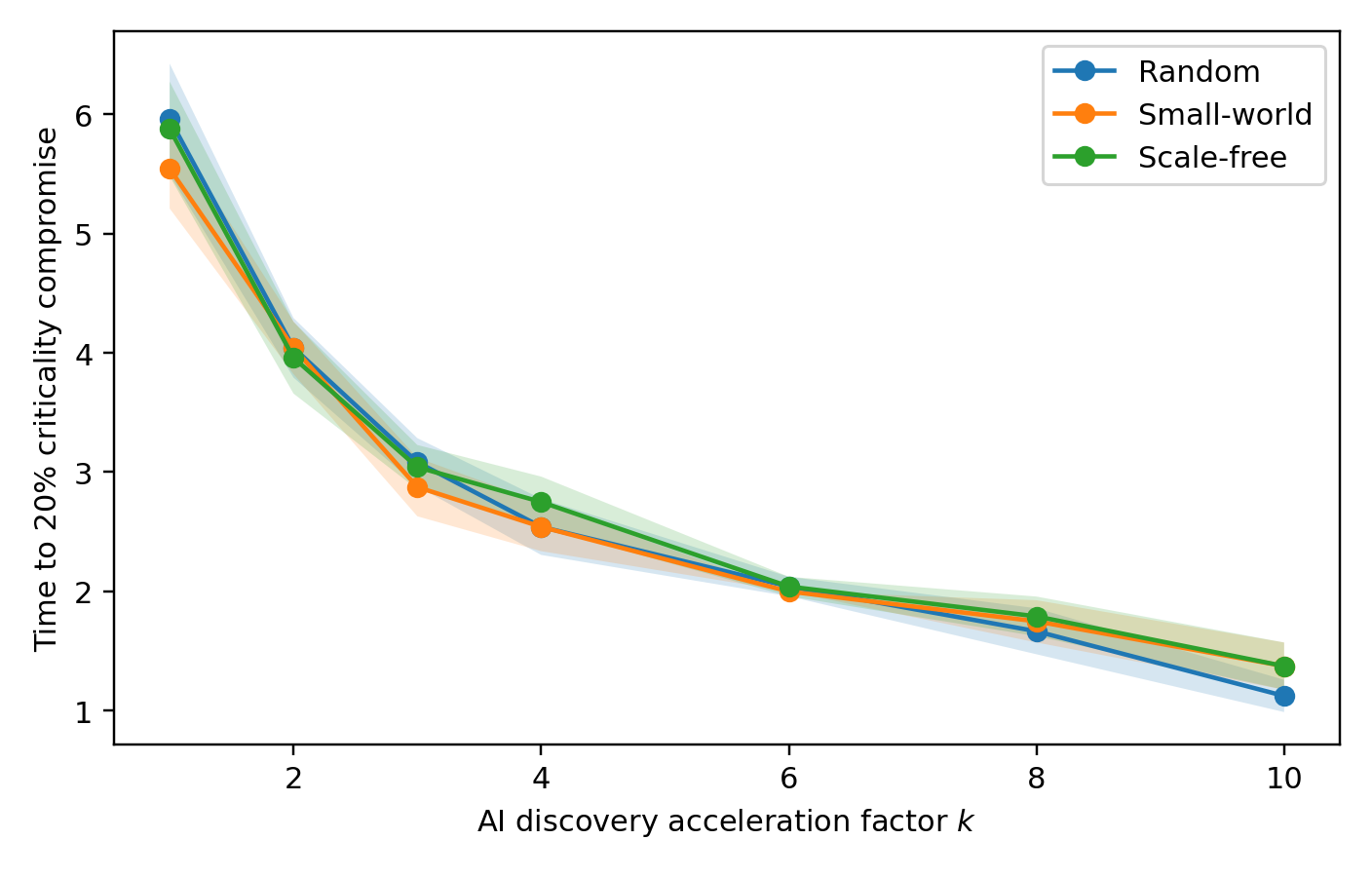}}
\caption{Mean time to 20\% criticality-weighted compromise. Shaded bands denote 95\% confidence intervals.}
\label{fig:ttc}
\end{figure}

Fig.~\ref{fig:ttc} shows that time to 20\% criticality-weighted compromise decreases as $k$ increases. This captures the operational consequence of discovery acceleration: defenders have less time to absorb, prioritize, and remediate vulnerability flows.

\subsection{Dynamic Coupling Sensitivity}
\begin{figure}[htbp]
\centerline{\includegraphics[width=0.48\textwidth]{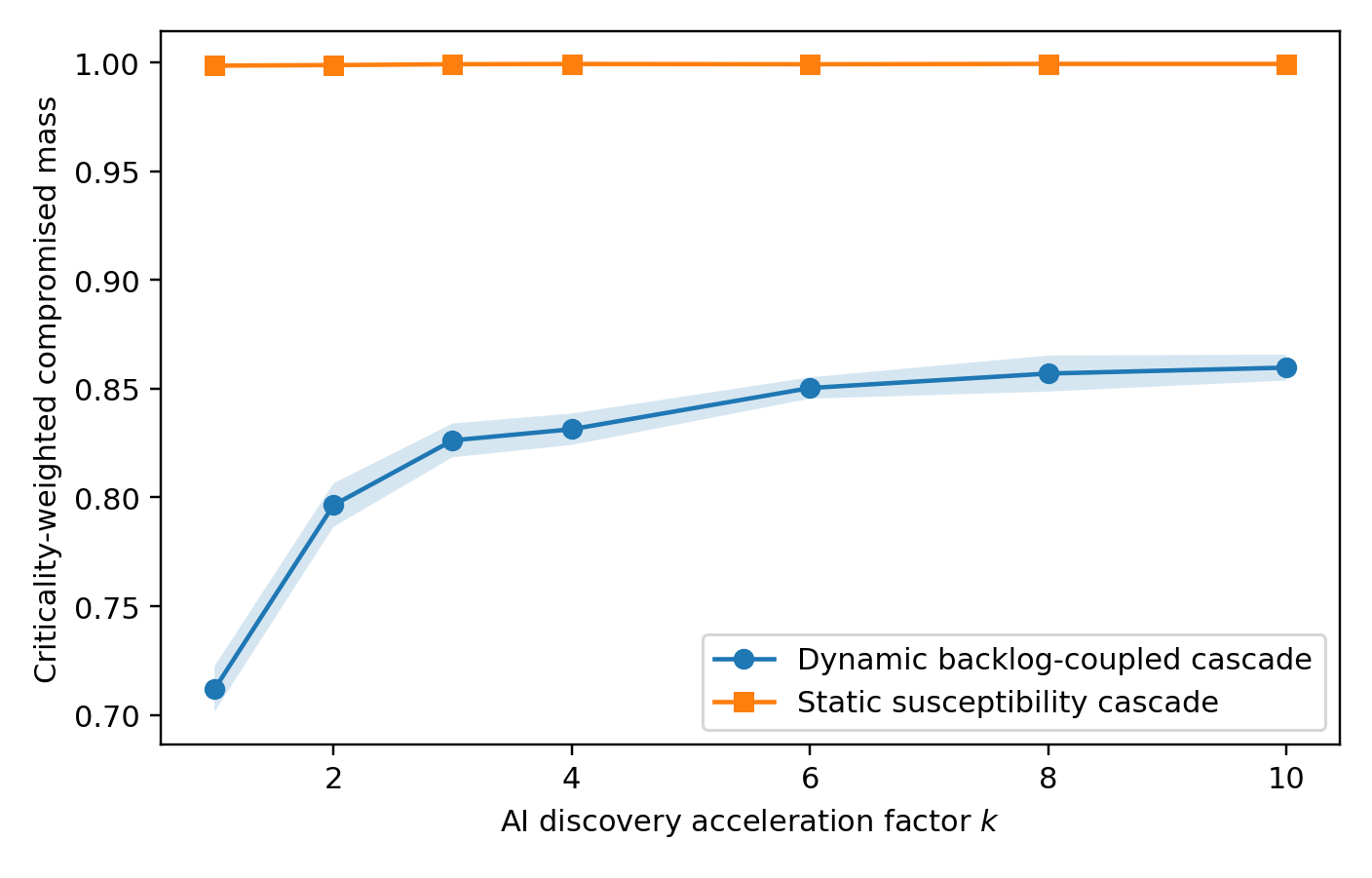}}
\caption{Dynamic backlog-coupled cascade versus static susceptibility cascade in scale-free topology. Shaded bands denote 95\% confidence intervals.}
\label{fig:dynamicstatic}
\end{figure}

To address whether the dynamic cascade model materially changes outcomes, Fig.~\ref{fig:dynamicstatic} compares the backlog-coupled cascade model to a static susceptibility model on the scale-free topology. The dynamic model produces different systemic risk behavior because unresolved backlog increases downstream susceptibility through $\psi_j(B_j(t))$. Shaded confidence bands are computed over repeated trials for both the dynamic and static variants. This confirms that the analytical modification is reflected in the simulation rather than merely asserted.

\subsection{Backlog Susceptibility Sensitivity}
\begin{figure}[htbp]
\centerline{\includegraphics[width=0.48\textwidth]{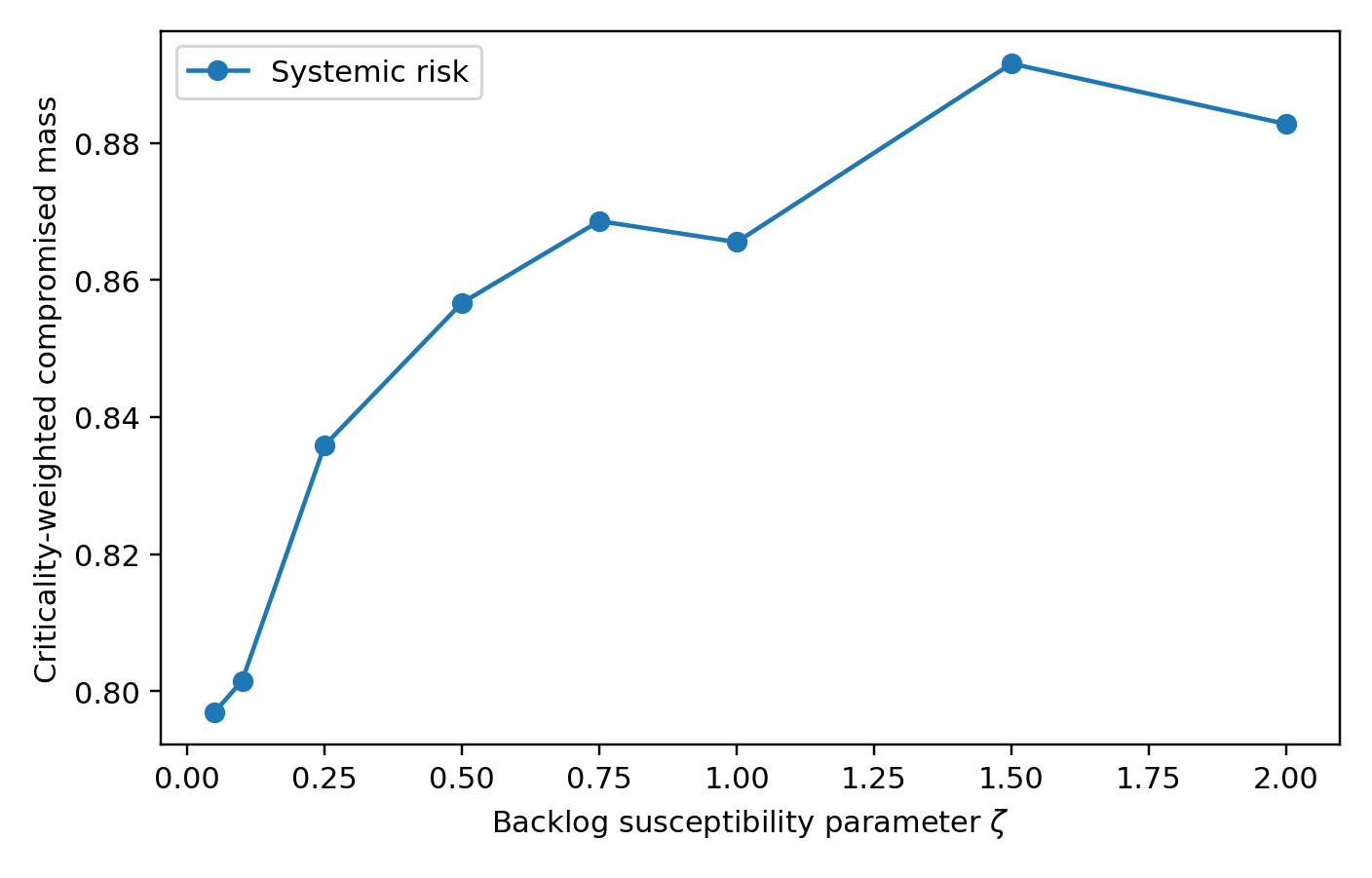}}
\caption{Sensitivity of systemic risk to backlog susceptibility parameter $\zeta$ in scale-free topology at $k=8$.}
\label{fig:zeta}
\end{figure}

Fig.~\ref{fig:zeta} varies the backlog susceptibility parameter $\zeta$ in $\psi_j(B_j)=1-\exp[-\zeta B_j]$ for the scale-free topology at $k=8$. Low $\zeta$ values mean that backlog weakly increases propagation susceptibility; high values mean that even small unresolved backlogs make downstream nodes susceptible. The curve shows that systemic risk is sensitive to $\zeta$ in the low-to-moderate range and then saturates as $\psi_j(B_j)$ approaches one.

\subsection{Intervention Analysis}
For a scale-free topology at $k=8$, we reran the intervention analysis under the current dynamic backlog-coupled model with $\zeta=0.55$ and 32 trials. The qualitative ordering is robust: remediation automation reduces backlog, segmentation reduces propagation, and the combined defense performs best overall.

\begin{table*}[t]
\caption{Intervention Results for Scale-Free Topology at $k=8$}
\centering
\small
\setlength{\tabcolsep}{4pt}
\begin{tabular}{lrrrr}
\toprule
Scenario & Backlog & Risk & LCC & TTC \\
\midrule
Baseline & 0.838$\pm$0.045 & 0.851$\pm$0.009 & 0.838$\pm$0.010 & 1.750$\pm$0.152 \\
Auto-remediation & 0.330$\pm$0.037 & 0.657$\pm$0.009 & 0.587$\pm$0.011 & 2.406$\pm$0.173 \\
Segmentation & 0.883$\pm$0.049 & 0.828$\pm$0.010 & 0.790$\pm$0.013 & 1.750$\pm$0.152 \\
Combined & 0.343$\pm$0.036 & 0.623$\pm$0.011 & 0.548$\pm$0.013 & 2.375$\pm$0.170 \\
\bottomrule
\end{tabular}
\label{tab:interventions}
\end{table*}

\begin{figure}[htbp]
\centerline{\includegraphics[width=0.48\textwidth]{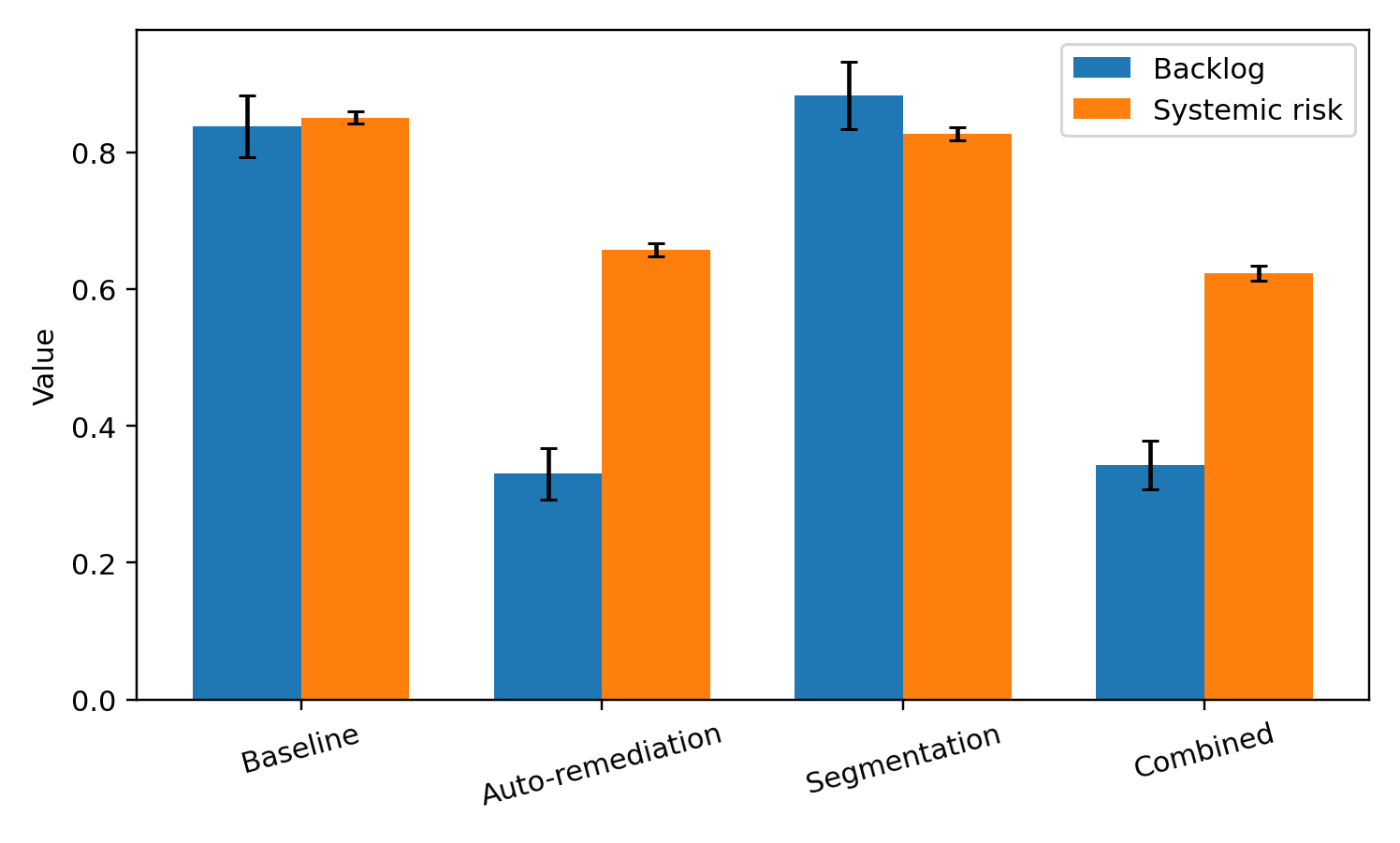}}
\caption{Intervention comparison under high discovery pressure in a scale-free topology. Error bars denote 95\% confidence intervals.}
\label{fig:interventions}
\end{figure}

Table~\ref{tab:interventions} summarizes the intervention results; the wide table is placed as a two-column float to avoid column overflow in IEEE format. These results support four conclusions. First, discovery acceleration can create remediation overload. Second, topology determines whether local exposure becomes systemic. Third, the dynamic backlog-coupled cascade model materially affects systemic risk estimates. Fourth, remediation automation and segmentation mitigate different mechanisms: remediation reduces local backlog, while segmentation reduces propagation.

\section{Discussion}
\subsection{Discovery Alone Is Insufficient}
The model suggests that organizations should not measure AI security success only by the number of vulnerabilities found. A tool that multiplies findings without increasing remediation capacity may increase overload. Useful metrics include validated findings remediated, exposure duration reduced, blast radius reduced, and criticality-weighted risk removed.

\subsection{Remediation Automation}
If AI increases discovery by factor $k$, stability requires either a corresponding increase in effective remediation capacity or a reduction in actionable arrivals. Automation can increase $\mu_i$ through patch generation, dependency updates, configuration fixes, test generation, deployment orchestration, and rollback mechanisms.

\subsection{Exploit-Aware Prioritization}
Severity-only prioritization is insufficient. Prioritization should account for exploitability, reachability, business criticality, graph centrality, control strength, and exploit availability. Under overload, triage quality becomes a central determinant of security.

\subsection{Segmentation and Attack-Surface Reduction}
Attack-surface reduction lowers discovery pressure by decreasing $A_i$. Segmentation lowers effective propagation weights by increasing $S_{ij}$. These interventions remain valuable even when remediation cannot keep up. The simulation suggests that in hub-dominated topologies, segmentation may reduce systemic risk more effectively than patch-speed improvements alone because it changes cascade structure rather than only reducing local backlog.

\subsection{Interpretation of the Intervention Result}
The intervention analysis should not be read as a universal claim that segmentation always dominates remediation. Rather, it identifies a regime: high discovery acceleration, hub-dominated topology, and strong propagation edges. In that regime, remediation automation lowers backlog, but systemic risk remains high unless graph coupling is reduced. The combined strategy is therefore the most robust.

\subsection{Operational Calibration Pathways}
Although the current study is synthetic, the model parameters can be estimated from operational telemetry. Discovery acceleration factors $k$ can be inferred from measured uplift in findings generated by AI-assisted testing relative to baseline workflows. Remediation capacities $\mu_i$ can be estimated from historical ticket closure rates or patch deployment throughput. Vulnerability introduction rates $\eta_i$ can be approximated from release cadence, dependency churn, and historical post-release defect rates. Edge weights $w_{ij}$ can be derived from identity trust relationships, network reachability, shared administrative domains, or service dependencies. Segmentation strengths $S_{ij}$ can be estimated from control assessments. Finally, the backlog susceptibility parameter $\zeta$ may be estimated by relating unresolved backlog states to observed incident propagation or near-miss data.

\subsection{Heterogeneous AI Acceleration}
The scalar acceleration factor $k$ is intended as a first-order control parameter rather than a claim of uniform AI impact. In practice, AI effectiveness will vary across languages, architectures, code maturity, test coverage, scanner integration, and vulnerability classes. A natural extension is a node-specific or time-varying process $k_i(t)$, allowing heterogeneous acceleration across the enterprise.

\subsection{Instantiation from Attack Graphs}
The weighted-edge enterprise graph need not be hand-specified. Existing attack-graph and reachability tools such as \cite{sheyner2002,noel2004,ou2006} can be used to derive edge candidates from privilege relationships, credential reuse paths, reachable services, firewall policy, identity federation, and lateral movement opportunities. In this setting, edge weights represent aggregated ease or likelihood of transition rather than arbitrary constants.

\subsection{Adversarial Co-Evolution}
The present model captures attacker-side AI effects indirectly through exploit-window compression $m_v$. A richer formulation would model attacker and defender adaptation jointly, where defenders allocate remediation resources while attackers prioritize targets, automate reconnaissance, or dynamically exploit newly surfaced weaknesses. Such attacker--defender co-evolution is an important direction for future work.

\subsection{Robustness of Triage Functional Form}
Equation (14) uses a parsimonious overload-response function for triage quality. The qualitative mechanism does not depend on this exact choice: alternative concave forms such as logistic saturation, exponential decay, or piecewise-linear overload penalties produce similar behavior in preliminary sensitivity checks. The essential requirement is that triage quality degrades when arrival load persistently exceeds handling capacity.

\subsection{Threshold Interpretation}
Theorem 1 clarifies the role of the cascade layer. Queue stability alone is insufficient: even a bounded backlog can be dangerous if the effective propagation matrix has spectral radius above the containment-adjusted threshold. Conversely, a large backlog may remain localized when segmentation reduces $\rho(M)$ enough to keep $\kappa<\kappa_c(B)$. This gives the paper a two-layer stability interpretation: remediation capacity governs backlog growth, while the spectral cascade threshold governs whether backlog-induced exposure becomes systemic.

This is consistent with prior findings that highly clustered communities can slow malicious propagation by temporarily containing infections within local neighborhoods \cite{xssworm2013}.

\section{Limitations and Future Work}
The model abstracts many real-world complexities. Vulnerabilities differ in exploitability, patch complexity, and attacker interest. Remediation capacity depends on people, tooling, vendor release cycles, testing pipelines, and business constraints. The propagation model abstracts exploit-chain details into weighted edges. AI acceleration factors are uncertain and may vary by codebase, language, architecture, and tool integration.

The simulation remains synthetic. Although this version updates the cascade implementation, reports confidence intervals, and reruns the intervention analysis under the dynamic backlog-coupled model, the results should be interpreted as qualitative validation of model behavior rather than empirical measurement of real-world compromise probabilities. Future work should calibrate parameters using enterprise vulnerability-management telemetry, public vulnerability datasets (e.g., NVD/CVE records), exploit intelligence feeds, incident reports, and dependency graphs. Even partial calibration from a single large organization would materially improve external validity.

The model could also be extended with priority queues, stochastic service times, adversarial behavior, and game-theoretic attacker-defender dynamics.

\section{Conclusion}
AI-accelerated vulnerability discovery may transform cybersecurity from a discovery-scarcity problem into a remediation-capacity and topology problem. When actionable vulnerability arrivals exceed remediation throughput, backlogs grow, triage quality degrades, exposure windows expand, and interconnected systems can amplify local weaknesses into systemic risk.

The strongest defense is not discovery alone. Resilient organizations must pair improved discovery with remediation automation, exploit-aware prioritization, attack-surface reduction, segmentation, and resilience planning for persistent backlog states.

\bibliographystyle{IEEEtran}
\bibliography{refs}

\end{document}